\documentclass[a4paper,12pt]{article}
\usepackage{graphicx}
\usepackage{cite}
\usepackage{amssymb,amsmath,latexsym,enumerate}
\usepackage{subfigure,epsfig}
\usepackage{array,longtable,lscape,bigints}

\begin{document}

\title{New non--standard Lagrangians for the Li\'{e}nard--type equations}
\author{Nikolai A. Kudryashov, Dmitry I. Sinelshchikov,\\
\texttt{nakudr@gmail.com}, \texttt{disine@gmail.com}, \vspace{0.5cm}\\
Department of Applied Mathematics, \\ National Research Nuclear University MEPhI, \\ 31 Kashirskoe Shosse, 115409 Moscow, Russian Federation}

\date{}

\maketitle

\begin{abstract}
Li\'{e}nard--type equations are used for the description of various phenomena in physics and other fields of science.
Here we find a new family of the Li\'{e}nard--type equations which admits a non--standard autonomous Lagrangian.
As a by-product we obtain autonomous first integrals for each member of this family of equations. We also show that some of the previously known conditions for the existence of a non--standard Lagrangian for the Li\'{e}nard--type equations follow from the linearizability of the corresponding equation via nonlocal transformations.
\end{abstract}

\noindent
Keywords: Li\'{e}nard--type equations; Lagrangian; inverse variational problem;

\section{Introduction}
We consider the following Li\'{e}nard--type equation
\begin{equation}
y_{zz}+f(y)y_{z}^{2}+g(y)y_{z}+h(y)=0,
\label{eq:L1}
\end{equation}
where $f$, $g$ and $h$ are arbitrary functions, which do not vanish simultaneously. We also suppose that $g(y)\neq 0$, that is we consider a dissipative case. Equation \eqref{eq:L1} appears in a wide range of applications such as physics, biology and mechanics \cite{Holmes,Polyanin}. Note that in the case of $f(y)=0$ equation \eqref{eq:L1} transforms to the classical Li\'{e}nard equation.

A problem of finding a Lagrangian for a given system of ordinary differential equations is an important problem of the calculus of variations (see, e.g. \cite{Lanczos}). Although the inverse variational problems for one--dimensional systems, such as equation \eqref{eq:L1}, can theoretically be solved (see, e.g. \cite{Jacobi,Whittaker,Bolza}), it seems difficult to find an explicit expression for the corresponding Lagrangian in the general case. Let us note that there are two types of Lagrangias: standard or natural ones and non--standard ones. If a Lagrangian is a quadratic form with respect to the velocities it is called a standard Lagrangian, otherwise a Lagrangian is called a non--standard one. Here we consider non--standard Lagrangians for equation \eqref{eq:L1}.

Several attempts have recently been made to find Lagrangians for the Li\'{e}nard--type equations. For example, in works \cite{Musielak2008,Musielak2009,Cieslinski2009} an ad hoc approach based on postulation of some form of a Lagrangian and finding the corresponding Euler--Lagrange equation was applied. As a result, authors found some families of Li\'{e}nard--type equations which admits both standard and non--standard Lagrangians. On the other hand, the Jacobi last multiplier leads to all possible Lagrangians for a one--dimensional system (see, e.g. \cite{Jacobi,Whittaker,Nucci2009,Nucci2010A}). For example, using this approach, several families of Li\'{e}nard--type equations that have a Lagrangian were found in \cite{Nucci2007,Nucci2010}. Authors of \cite{Nucci2009} constructed Lagrangians for many of the Painlev\'{e}--Gambi\'{e}r equations  by means of the Jacobi last multiplier. Results obtained in \cite{Musielak2008,Musielak2009,Cieslinski2009} were rederived via the Jacobi approach in \cite{Lakshmanan2016}.

In this work we extend the previously obtained results and find a new family of the Li\'{e}nard--type equations that admits variation formulation and the corresponding family of Lagrangians. To this end we use an approach based on application of nonlocal transformations \cite{Kudryashov2015,Kudryashov2016}. First, we find that a particular equation from family \eqref{eq:L1} admits a new autonomous non--standard Lagrangian. Then we obtain a correlation on functions $f$, $g$ and $h$ that provides a correspondence via nonlocal transformations between this equation and equation \eqref{eq:L1}. As a result, we find a new family of the Li\'{e}nard--type equations that has a non--standard autonomous Lagrangian. We also obtain autonomous first integrals for each member of this equations family. Moreover, we show that some of the conditions for the existence of non--standard Lagrangian for equation \eqref{eq:L1} that were found in \cite{Musielak2008,Musielak2009,Cieslinski2009,Lakshmanan2016} follows from linearizability via nonlocal transformations of the corresponding family of Li\'{e}nard--type equations. To the best of our knowledge, our results have not been reported previously.

The rest of this work is organized as follows. In the next section we give a new family of the Li\'{e}nard--type equations that admits an autonomous non--standard Lagrangian and discuss some previously obtained conditions for the existence of a Lagrangian for equation \eqref{eq:L1}. In Section 3 we present two new examples of Li\'{e}nard--type equations that admit autonomous Lagrangians and first integrals. In the last section we briefly summarize our results.

\section{Main results}

In this section we study equation \eqref{eq:L1} with the help of the following nonlocal transformations
\begin{equation}
w=F(y), \quad d\zeta=G(y)dz, \quad F_{y}G\neq0,
\label{eq:L1_1}
\end{equation}
where $\zeta$ and $w$ are new independent and dependent variables correspondingly. Transformations \eqref{eq:L1_1} are often called the generalized Sundman transformations (see, e.g. \cite{Duarte1994,Meleshko2010}).

First of all, we demonstrate that equation \eqref{eq:L1} can be transformed  into the Li\'{e}nard equation via \eqref{eq:L1_1}.
Indeed, substituting \eqref{eq:L1_1} with $F(y)=y$ and $G(y)=\exp\{-\int f dy\}$ into \eqref{eq:L1} we obtain $y_{\zeta\zeta}+\tilde{g}(y)y_{\zeta}+\tilde{h}(y)=0$, where $\tilde{g}(y)=\exp\{\int f dy\}g(y)$ and $\tilde{h}(y)=\exp\{2\int f dy\}h(y)$. Since a combination of nonlocal transformations is a nonlocal transformation, all results obtained for the Li\'{e}nard equation can be extended to Li\'{e}nard--type equation \eqref{eq:L1}. Thus, without loss of generality, further we assume that $f(y)=0$.  In order to obtain results for equation \eqref{eq:L1} we have to make the following substitutions
\begin{equation}
g(y)\rightarrow e^{\int f dy}g(y), \quad h(y) \rightarrow e^{2\int f dy}h(y), \quad y_{z} \rightarrow e^{\int f dy} y_{z} \,.
\label{eq:L1_3_1}
\end{equation}

Now we discuss linearizabily of equation \eqref{eq:L1} by means of transformations \eqref{eq:L1_1} and the corresponding Lagrangians. We recall that the following damped harmonic oscillator
\begin{equation}
w_{\zeta\zeta}+\sigma w_{\zeta}+(2\sigma^{2}/9)w=0,
\label{eq:L1_3}
\end{equation}
where $\sigma\neq0$ is an arbitrary parameter, admits the following non--standard Lagrangian \cite{Musielak2008,Musielak2009}
\begin{equation}
%L=\frac{1}{w_{\zeta}+2\sigma/3w}.
L=[w_{\zeta}+(2\sigma/3)w]^{-1}.
\label{eq:L1_5}
\end{equation}
Let us apply  criterion of equivalence between \eqref{eq:L1} with $f(y)=0$ and \eqref{eq:L1_3} via \eqref{eq:L1_1} which was obtained in \cite{Kudryashov2016} (see also \cite{Meleshko2010}).

\textbf{Proposition 1} Equation \eqref{eq:L1} can be transformed into \eqref{eq:L1_3} by means of \eqref{eq:L1_1} with $
F(y)=\mathcal{G}$ , $G(y)=(1/\sigma) g(y)$ if the following correlation holds
\begin{equation}
h=(2/9)\,g(y)\,\mathcal{G}.
\label{eq:L1_9}
\end{equation}
Here and below we denote by $\mathcal{G}$
\begin{equation}
\mathcal{G}=\left.\int\right. g(y)dy+\kappa,
\label{eq:L1_7_1}
\end{equation}
where $\kappa$ is an arbitrary parameter.

Criterion of linearizability \eqref{eq:L1_9} is completely equivalent to the criterion for the existence of a non--standard Lagrangian obtained in \cite{Musielak2008,Musielak2009}. Indeed, assuming that $\sigma=1$ and $\kappa=0$ and changing notation to the one of works \cite{Musielak2008,Musielak2009} (see also \cite{Cieslinski2009,Lakshmanan2016}) and differentiate the result with respect to $y$ we get the criterion of the existence of a non--standard Lagrangian obtained in \cite{Musielak2008,Musielak2009}. We get the corresponding Lagrangian as follows
\begin{equation}
%L=\frac{1}{\sigma y_{z}+(2\sigma/3)\mathcal{G}}.
L=[\sigma y_{z}+(2\sigma/3)\mathcal{G}]^{-1}.
\label{eq:L1_9_1}
\end{equation}
Taking into account \eqref{eq:L1_3_1}, from \eqref{eq:L1_9_1} we obtain a non--standard Lagrangian for equation \eqref{eq:L1} with $f(y)\neq0$
\begin{equation}
%L=\frac{1}{\sigma e^{\int f dy} y_{z}+2\sigma/3(\int e^{\int f dy} g(y) dy +\kappa)},
L=\left[\sigma e^{\int f dy} y_{z}+(2\sigma/3)\left(\int e^{\int f dy} g(y) dy +\kappa\right)\right]^{-1},
\label{eq:L1_9_3}
\end{equation}
at the same time correlation \eqref{eq:L1_9} becomes
\begin{equation}
h=\frac{2}{9}g(y)e^{-\int f dy}\left(\int e^{\int f dy} g(y)dy+\kappa\right).
\label{eq:L1_9_5}
\end{equation}
The same Lagrangian as \eqref{eq:L1_9_3} was obtained in \cite{Musielak2008,Musielak2009,Cieslinski2009,Lakshmanan2016}. Therefore, we have seen that the family of Li\'{e}nard--type equations \eqref{eq:L1} defined by correlation \eqref{eq:L1_9_5} shares Lagrangian \eqref{eq:L1_5} with \eqref{eq:L1_3}.

Now we find a new family of the Li\'{e}nard--type equations which admits a non--standard Lagrangian. Let us consider a sub--case of equation \eqref{eq:L1} that is of the Painlev\'{e}--Gambi\'{e}r type (see \cite{Ince}, equation XXVIII at $q=0$) and has the form
\begin{equation}
ww_{\zeta\zeta}-\frac{1}{2}w_{\zeta}^{2}+w^{2}w_{\zeta}-\frac{1}{2}w^{4}+72H=0,
\label{eq:Ince_XXVIII}
\end{equation}
where $H\neq0$ is an arbitrary parameter. Notice that at $H=0$ equation \eqref{eq:Ince_XXVIII} can be linearized via \eqref{eq:L1_1}, and, thus, we do not consider this case.

One can show that the corresponding partial differential equation for the Lagrangian of \eqref{eq:Ince_XXVIII} (see, e.g. \cite{Whittaker,Bolza}) admits the following solution in the autonomous case:
\begin{equation}
L=-(1/3)(w_{\zeta}+w^{2})^{2}w^{-2}\Bigg[(w_{\zeta}+w^{2})^{2}-864H\Bigg]+20736H^{2}w^{-2}.
\label{eq:Lagrangian}
\end{equation}
Indeed, substituting \eqref{eq:Lagrangian} into the Euler--Lagrange equation $(L_{w_{\zeta}})_{\zeta}-L_{w}=0$ we get \eqref{eq:Ince_XXVIII}. Note that, actually, we have found an infinite family of Lagrangians, since we can add to \eqref{eq:Lagrangian} any function which is a total derivative of a differentiable function. We can also find the Jacobi last multiplier for equation \eqref{eq:Ince_XXVIII} either using a connection between a Lagrangian and the Jacobi last multiplier or finding directly a solution of the partial differential equation for the Jacobi last multiplier (see \cite{Whittaker,Nucci2009}). As a result, we get
\begin{equation}
M=4[144H-(w_{\zeta}+w^{2})^{2}]w^{-2}.
\label{eq:JLM}
\end{equation}
Calculating the canonical momentum $p$ as $ L_{w_{\zeta}}$ we find an autonomous first integral of equation \eqref{eq:Ince_XXVIII} using the relation $I=p w_{\zeta}-L$:
\begin{equation}
I=-(1/3)(3w_{\zeta}-w^{2})(w_{\zeta}+w^{2})^{3}\,w^{-2}+288H(w_{\zeta}^{2}-w^{4}-72H)w^{-2}.
\label{eq:L3_5}
\end{equation}

Let us study the case when the whole family of equations \eqref{eq:L1} can be transformed into \eqref{eq:Ince_XXVIII} via \eqref{eq:L1_1}.

\textbf{Theorem 1.} Equation \eqref{eq:L1} with $f(y)=0$ can be transformed into \eqref{eq:Ince_XXVIII} with the help of transformations \eqref{eq:L1_1} with
\begin{equation}
F(y)=\lambda \mathcal{G}^{2/3}, \quad G=g(y) \lambda ^{-1}\mathcal{G}^{-2/3},
\label{eq:L3_7}
\end{equation}
if the following correlation holds
\begin{equation}
h(y)=(3/4)g(y)\lambda^{-4}\mathcal{G}^{-5/3}\left[144H-\lambda^{4}\mathcal{G}^{8/3} \right],
\label{eq:L3_9}
\end{equation}
where $\lambda\neq0$ is an arbitrary parameter and $\mathcal{G}$ is defined by \eqref{eq:L1_7_1}.

\textbf{Proof}. We calculate $y_{z}$, $y_{zz}$ via $w_{\zeta}$, $w_{\zeta\zeta}$ and then substitute the result into \eqref{eq:L1} with $f(y)=0$. As a consequence, we get an ordinary differential  equation for $w$. Requiring that this equation is \eqref{eq:Ince_XXVIII} we obtain formulas \eqref{eq:L3_7} and correlation \eqref{eq:L3_9}. This completes the proof.

\textbf{Corollary 1}. Equation \eqref{eq:L1} in the case of $f(y)=0$ and when $h(y)$ satisfies \eqref{eq:L3_9} admits the following non--standard Lagrangian
\begin{equation}
\begin{gathered}
L=-(\lambda^{2}/3)\left[\frac{2}{3}y_{z}+ \mathcal{G}\right]^{2} \mathcal{G}^{-2/3}
\left[ \mathcal{G}^{2/3}\lambda^{4}\left[\frac{2}{3}y_{z}+ \mathcal{G}\right]^{2}-864H\right]
+\frac{20736H^{2}}{\lambda^{2}\mathcal{G}^{4/3}}.
\label{eq:L3_11}
\end{gathered}
\end{equation}
\textbf{Corollary 2}. An autonomous first integral for equation \eqref{eq:L1} satisfying correlation \eqref{eq:L3_9} and at $f(y)=0$ has the form
\begin{equation}
\begin{gathered}
I=-\frac{\lambda^{6}\left(2y_{z}-\mathcal{G}\right)}{3}\left(\frac{2}{3}y_{z}+\mathcal{G}\right)^{3}+\frac{288H\lambda^{2}}{\mathcal{G}^{2/3}}\left(\frac{4}{9}y_{z}^{2}-
\mathcal{G}^{2}\right)-\frac{20736H^{2}}{\lambda^{2}\mathcal{G}^{4/3}}.
\label{eq:L3_11_1}
\end{gathered}
\end{equation}

One can see that condition \eqref{eq:L3_9} does not coincide with conditions for the existence of a Lagrangian for \eqref{eq:L1} that were found in \cite{Musielak2008,Musielak2009,Cieslinski2009,Lakshmanan2016}. Therefore, we obtain a new family of the Li\'{e}nard--type equations that admits a non--standard Lagrangian. We can also obtain the Jacobi last multiplier for every member of this family of equations employing \eqref{eq:JLM} and transformations \eqref{eq:L1_1} with \eqref{eq:L3_7}.  Notice also that one can find the general solution of equation \eqref{eq:L1} in the case of $f(y)=0$ and when $h(y)$ satisfies \eqref{eq:L3_9} using transformations \eqref{eq:L1_1} with \eqref{eq:L3_7} and the general solution of \eqref{eq:Ince_XXVIII}. Detailed analysis of this solution will be reported elsewhere. Moreover, note that transformations \eqref{eq:L1_1} with \eqref{eq:L3_7} do not coincide with those presented in \cite{Euler}. Finally, let us remark that one can obtain a Lagrangian, Jacobi last multiplier  and first integral for equation \eqref{eq:L1} when $f(y)\neq0$ along with the corresponding criterion for their existence with the help of \eqref{eq:L1_3_1}.

\section{Examples}

In this section we consider applications of Theorem 1 and provide new examples of equations from family \eqref{eq:L1} with non--standard Lagrangians.

\textit{Example 1.} Let us suppose that $f(y)=0$, $g(y)=2y^{2}$ and $\lambda=2^{1/3}3^{7/6}$. Then from \eqref{eq:L1} and \eqref{eq:L3_9} we obtain the corresponding Li\'{e}nard equation
\begin{equation}
y_{zz}+2y^{2}y_{z}-y^{5}+Hy^{-3}=0.
\label{eq:L5_1}
\end{equation}
By means of Corollary 1 we get a non--standard Lagrangian for equation \eqref{eq:L5_1}
\begin{equation}
L=-576\left(\left[(y_{z}+y^{3})^{2}-3Hy^{-2}\right]^{2}-12H^{2}y^{-4}\right)
\label{eq:L5_3}
\end{equation}
Using Corollary 2 we find an autonomous first integral for equation \eqref{eq:L5_1}
\begin{equation}
I=-576(3y_{z}-y^{3})(y_{z}+y^{3})^{3}+1728H(2y^{2}y_{z}^{2}-2y^{8}-H)y^{-4}.
\label{eq:L5_5}
\end{equation}
Thus, we see that damped anharmonic oscillator \eqref{eq:L5_1} has both autonomous non--standard Lagrangian \eqref{eq:L5_3} and  autonomous first integral \eqref{eq:L5_5}.

\textit{Example 2.} Now we consider an example of equation \eqref{eq:L1} with $f(y)\neq0$. We assume that $f(y)=-7/(2y)$, $g(y)=-\beta y$, $\kappa=0$ and $\lambda=2^{1/12}3^{7/6}\beta^{-1/6}$, where $\beta\neq0$ is an arbitrary parameter. Then from \eqref{eq:L1}, \eqref{eq:L1_3_1} and \eqref{eq:L3_9}  we get
\begin{equation}
y_{zz}-\frac{7}{2y}y_{z}^{2}-\beta y y_{z}+\frac{\beta^{2}}{2}y^{3}-Hy^{7}=0
\label{eq:L5_7}
\end{equation}
With the help of Corollary 1 and \eqref{eq:L1_3_1} we find an autonomous Lagrangian for equation \eqref{eq:L5_7}
\begin{equation}
L=-144\sqrt{2}\,\beta^{-1}\left[{(y_{z}+\beta y^{2})^{4}}{y^{-14}}-{12H(y_{z}+\beta y^{2})^{2}}{y^{-6}}-12H^{2}y^{2}\right]
\label{eq:L5_9}
\end{equation}
Using Corollary 2 and \eqref{eq:L1_3_1} we get a first integral for equation \eqref{eq:L5_7}
\begin{equation}
I=-{144\sqrt{2}}\beta^{-1} y^{-14}\left[(3y_{z}-\beta y^{2})(y_{z}+\beta y^{2})^{3}-12Hy^{8}(y_{z}^{2}-\beta^{2}y^{4})+12H^{2}y^{16}\right]
\label{eq:L5_11}
\end{equation}
Therefore, we see that equation \eqref{eq:L5_7} admits both autonomous Lagrangian and first integral. It is worth noting that equation \eqref{eq:L5_7} can be considered as the traveling--wave reduction of the generalized Fisher equation (see, e.g. \cite{Kudryashov2014a}).

\section{Conclusion}
In this work we have considered Li\'{e}nard--type equation \eqref{eq:L1}. We have found a new family of the Li\'{e}nard--type equations which admits a non--standard Lagrangian. Using this Lagrangian we have obtained autonomous first integrals for each member of this family of equations. We have illustrated our results by providing two new examples of Li\'{e}nard--type equation that admit both a non--standard Lagrangian and an autonomous first integral. We have also shown that some of the previously reported non--standard Lagrangians of this equation can be obtained from a non--standard Lagrangian for a special case of the damped harmonic oscillator.

\section{Acknowledgments}
Authors are grateful for an anonymous reviewer's valuable comments and suggestions. This research was partially supported by grant for the state support of young Russian scientists 6624.2016.1, by RFBR grant 14--01--00498 and the Competitiveness Program of NRNU 'MEPhI'.

\end{document}